\documentclass[prl,twocolumn,showpacs,superscriptaddress,nofootinbib,floatfix,10pt]{revtex4-1} %notitlepage,
\usepackage[utf8]{inputenc}
\usepackage{color}
\usepackage{amsfonts,amsmath,amssymb}
\usepackage{longtable,booktabs}
\usepackage{graphicx}
\usepackage{bm}
\usepackage[colorlinks=true,linkcolor=blue,citecolor=blue]{hyperref}
\usepackage{color}
\usepackage{epstopdf}
\usepackage{mathtools}
\usepackage{subfigure}
\newcommand{\ddstaro}{D^0\bar D^{*0}}

\newcommand{\X}{X(3872)}

\newcommand{\C}{\mathcal{C}}
\newcommand{\A}{\mathcal{A}}
\newcommand{\B}{\mathcal{B}}
\newcommand{\order}[1]{{\mathcal{O}\!\left(#1\right)}}

\begin{document}

\title{$D^{\pm}D^{*\mp}$ Hadronic Atom as a Key to Revealing the $X(3872)$ Mystery}

\author{Zhen-Hua Zhang}\email{zhangzhenhua@itp.ac.cn}
\author{Feng-Kun Guo}\email[Corresponding author, ]{fkguo@itp.ac.cn}

\affiliation{CAS Key Laboratory of Theoretical Physics, Institute of Theoretical Physics, Chinese Academy of Sciences, Beijing 100190, China}
\affiliation{School of Physical Sciences, University of Chinese Academy of Sciences, Beijing 100049, China
}

\begin{abstract}

The $X(3872)$, whose mass coincides with the $D^0\bar D^{*0}$ threshold, is the most extended hadron object.
Since its discovery in 2003, debates have never stopped regarding its internal structure. 
We propose a new object, the $X$ atom, which is the $D^\pm D^{*\mp}$ composite system with positive charge parity and a mass of $(3879.89\pm0.07)$~MeV, formed mainly due to the Coulomb force. We show that a null signal of the $X$ atom can be used to put a lower limit on the binding energy of the $X(3872)$. From the current knowledge of the $X(3872)$ properties, the production rate for the $X$ atom relative to the $X(3872)$ in $B$ decays and at hadron colliders should be at least $1\times10^{-3}$. New insights into the $X(3872)$ will be obtained through studying the $X$ atom.

\end{abstract}

\date{\today}

\maketitle

\vspace{2cm}

{\it Introduction.}---Hadron resonances containing a pair of charm and anticharm quarks are being intensively studied at ongoing and planned high energy experiments, such as BESIII~\cite{Ablikim:2019hff}, LHCb~\cite{Cerri:2018ypt}, Belle-II~\cite{Kou:2018nap}, PANDA~\cite{Lutz:2009ff}, and so on.
The physics motivation is to understand the tens of mysterious hadron resonances in the charmonium mass regime, called the $XYZ$ states, that have properties beyond theoretical expectations.
Among them, the $X(3872)$, also known as $\chi_{c1}(3872)$~\cite{Zyla:2020zbs}, is the most mysterious one. Since its discovery in 2003 by the Belle Collaboration~\cite{Choi:2003ue}, debates regarding its internal structure have never stopped (for a few recent reviews, see Refs.~\cite{Chen:2016qju,Hosaka:2016pey,Lebed:2016hpi,Esposito:2016noz,Guo:2017jvc,Ali:2017jda,Olsen:2017bmm,Kalashnikova:2018vkv,Liu:2019zoy,Brambilla:2019esw}).

The most salient feature of the $\X$ is that its mass coincides exactly with the $\ddstaro$ threshold, with a difference (to be called binding energy),
\begin{equation}
    \delta = m_{D^0} + m_{D^{*0}} - m_X = (0.00\pm0.18)~\text{MeV}, 
    \label{eq:delta}
\end{equation} 
where we have used the ``OUR AVERAGE'' values for the masses: $m_{D^0}=(1864.84\pm0.05)$~MeV, $m_{D^{*0}}=(2006.85\pm0.15)$~MeV, and $m_{X}=(3871.69\pm0.17)$~MeV in the \textit{Review of Particle Physics}~\cite{Zyla:2020zbs} [for recent LHCb measurements of the $\X$ resonance parameters, see Refs.~\cite{Aaij:2020qga,Aaij:2020xjx} ].
Despite the closeness to the threshold, the $\X$ couples strongly to the $\ddstaro$ channel, which is manifested by the large branching fraction to the $D^0\bar D^{*0}$ and $D^0\bar D^0\pi^0$ channels~\cite{Zyla:2020zbs,Li:2019kpj,Braaten:2019ags}. 

From the uncertainty principle, the closeness of the $\X$ to the $\ddstaro$ threshold and the strong coupling indicate that the wave function of the $\X$ at long distances is given by that of the $\ddstaro$ component, which has a size of $r_X \simeq (2\mu_0|\delta|)^{-1/2}\gtrsim 10$~fm, where $\mu_0$ is the $D^0\bar D^{*0}$ reduced mass. The typical scale for the relative momentum between the neutral charmed mesons, $\sqrt{2\mu_0|\delta|} \lesssim 19$~MeV, is much smaller than strong and weak interaction scales, leading to a factorization between the short and long-distance contributions to the productions of the $\X$ at high-energy hadron colliders and in $B$ decays~\cite{Braaten:2005jj}.
The long-distance part, which can be computed in terms of nonrelativistic effective field theory (NREFT), is universal and depends only on the binding energy.
The production rates of the $\X$ are then proportional to the absolute square of the universal transition amplitude (or the effective coupling constant) from the $\ddstaro$ meson pair to the $\X$, which is proportional to $\sqrt{\delta}$~\cite{Braaten:2004rw,Braaten:2005jj,Artoisenet:2010va,Guo:2014sca,Guo:2017jvc},
\begin{equation}
    g_X^2 = \frac{2\pi}{\mu_0^2} \sqrt{2\mu_0 \delta} .
    \label{eq:gx}
\end{equation}
Thus, in line with intuition, the production of very loosely bound states is suppressed.
For debates related to whether the production of the $\X$ at hadron colliders can be used to discriminate possible internal structures of the $\X$, see Refs.~\cite{Bignamini:2009sk,Artoisenet:2009wk,Meng:2013gga,Guo:2014sca,Albaladejo:2017blx,Guo:2017jvc,Esposito:2017qef,Wang:2017gay,Braaten:2018eov,Butenschoen:2019npa,Zhang:2020dwn,Esposito:2020ywk}.

In order to understand the nature of the $\X$, it is important to have precise measurements of its binding energy and decay width, since these quantities are closely related to the long-distance, and thus the hadronic molecular, component of the $\X$.
The most precise measurement was given by the Flatt\'e analysis of the $\X$ events in the $J/\psi\pi^+\pi^-$ decay mode from bottom-hadron decays collected at the LHCb experiment~\cite{Aaij:2020qga}. The PANDA experiment under construction is able to measure the width and line shape of the $\X$ with a precision of 100~keV~\cite{PANDA:2018zjt}, and it is also able to measure the $\X$ binding energy with a precision well beyond Eq.~\eqref{eq:delta}~\cite{Sakai:2020crh} with the method proposed in Ref.~\cite{Guo:2019qcn}. 
One notices that the pole of the $\X$ in the LHCb analysis is below the $\ddstaro$ threshold in most of the confidence region~\cite{Aaij:2020qga}.

In this Letter, we propose to investigate a new object, the $D^\pm D^{*\mp}$ atom, to be called the $X$ atom below. Its production can be related to that of the $\X$ in a model-independent way using NREFT. Consequently, the null signal so far in the high quality LHCb data~\cite{Aaij:2020qga,Aaij:2020xjx} can be used to set a lower bound of the $\X$ binding energy.

{\it Hadronic atom.}---One notices that the radius of the $\X$, $r_X\gtrsim10$~fm, implies that the $\ddstaro$ component of the wave function is extremely extended from the point of view of strong interactions.
It even has the same order of magnitude as the Bohr radius of a hadronic atom formed of two oppositely charged charmed mesons, which are bound together mainly because of the electric Coulomb force.
The Bohr radius for the Coulomb bound state of $D^\pm D^{*\mp}$ is 
\begin{equation}
    r_c = \frac1{\alpha \mu_c} = 27.86~\text{fm},
    \label{eq:rc}
\end{equation}
where $\alpha$ is the fine structure constant, and $\mu_c$ is the $D^+D^{*-}$ reduced mass.
The Coulomb binding energy is given by 
\begin{equation}
    E_n = \frac{\alpha^2 \mu_c}{2n^2},
\end{equation}
with $n$ the principal quantum number. For the ground state, $n=1$, one has $E_1 = 25.81$~keV.
This value is within the uncertainty for the $\X$ binding energy in Eq.~\eqref{eq:delta} and also in the range of the updated LHCb analysis~\cite{Aaij:2020qga}, meaning that the size of the $X$ atom is comparable with that of the $\X$.

When two hadrons are well separated at a distance of the Bohr radius scale, much larger than the typical strong interaction radius, $1/\Lambda_\text{QCD}$ with $\Lambda_\text{QCD} = \mathcal{O}(300~\text{MeV})$,
the strong interaction effect must be weak, and can be treated as a correction to the dominant Coulomb force.
Furthermore, the system receives only influences of the strong interaction at the longest distance, and thus probes the strong interaction strength at threshold.
So far, hadronic atoms have only been studied for systems made of light hadrons, such as charged pions, kaons, and the proton, and have been used to extract the scattering lengths in such systems (for reviews, we refer to Refs.~\cite{Gasser:2007zt,Gasser:2009wf}). 
The $X$ atom is different from these hadronic atoms because the strong interaction in this case is nonperturbative due to the existence of the $\X$ close to the $\ddstaro$ threshold, which is 8.2~MeV below the $D^+D^{*-}$ threshold (and the $X$ atom). Therefore, the strong interaction correction to the Coulomb force needs to be treated in a nonperturbative way.

Let us start with the Lagrangian for the $\ddstaro$ and $D^+D^{*-}$ (the charge conjugated channels are implicit) coupled-channel system with positive charge parity, which is relevant for the $\X$:
\begin{align}
    \mathcal{L}&=-\frac{1}{4}F_{\mu\nu}F^{\mu\nu}+\sum_{\phi=D^\pm,D^0,\bar D^0}\phi^{\dagger}\left(iD_t-m_{\phi}+\frac{\nabla^2}{2m_{\phi}}\right)\phi \notag\\ 
    &+\sum_{\phi=D^{*\pm},D^{*0},\bar D^{*0}} \phi^{\dagger}\left(iD_t-m_\phi+i\frac{\Gamma_\phi}{2}+\frac{\nabla^2}{2m_\phi}\right)\phi \notag\\
    &-\frac{C_{0}}{2}(D^{+}D^{*-}-D^{-}D^{*+})^{\dagger}(D^{+}D^{*-}-D^{-}D^{*+}) \notag\\ 
    &-\frac{C_{0}}{2}\left[(D^{+}D^{*-}-D^{-}D^{*+})^{\dagger}(D^{0}\bar{D}^{*0}-\bar{D}^{0}D^{*0})+\mathrm{H.c.}\right] \notag\\
    &-\frac{C_{0}}{2}(D^{0}\bar{D}^{*0}-\bar{D}^{0}D^{*0})^{\dagger}(D^{0}\bar{D}^{*0}-\bar{D}^{0}D^{*0})+\cdots,
    \label{eq:lag}
\end{align}
where $F_{\mu\nu}=\partial_\mu A_\nu - \partial_\nu A_\mu$ is the electromagnetic field strength tensor, $D_t \phi = \partial_t \phi \mp i Q A_0 \phi$ with $Q$ the electric charge of the $\phi$ field, and $C_0$ is the constant contact term parametrizing the strong interaction between the charmed mesons. The phase convention for the charge conjugation is chosen such that $\C D \C^{-1} = \bar D$ and $\C D^* \C^{-1} = - \bar D^*$ with $\C$ the charge conjugation operator.
We have neglected the isospin-vector contribution for the strong interaction part, which should be a good approximation because there is an $\X$ close to the $\ddstaro$ threshold, while there is no isovector state with positive charge parity~\cite{Gamermann:2009uq}.
Although the ratio of branching fractions of the $X(3872)$ decays into the isovector $J/\psi\rho$ and isoscalar $J/\psi\omega$ exhibits a huge isospin breaking, it is shown in Ref.~\cite{Hanhart:2011tn} that the isospin breaking comes mainly from the difference in phase spaces, and the effective coupling of the $X(3872)$ to $J/\psi\rho$ is much smaller than that to $J/\psi\omega$, with a ratio $0.26^{+0.08}_{-0.05}$. Thus, we expect that the approximation neglecting the isospin-vector part should work at the level of 30\%. With such an approximation, the following expressions can be written in a more compact form.
Terms for the relativistic corrections and higher order electromagnetic corrections have been neglected since their contribution to the binding energy starts from $\order{\alpha^4}$.

We have introduced the widths of the $D^*$ mesons as constants into the Lagrangian, as done in, e.g., Ref.~\cite{Braaten:2020nmc}. 
It has been shown in Ref.~\cite{Hanhart:2010wh} that this is a very good approximation when $\Gamma_\phi/(2E_\phi)\ll 1$. Here, $E_\phi$ is the difference between the mass of $D^*$ and the threshold of $D\pi$ that the $D^*$ decays into; $E_\phi$ is about 7~MeV for $D^{*0}$ and about 6~MeV for $D^{*+}$, much larger than the $D^*$ widths: $\Gamma_{D^{*+}} = (83.4\pm1.8)$~keV~\cite{Zyla:2020zbs} and $\Gamma_{D^{*0}}=(55.3\pm1.4)$~keV~\cite{Guo:2019qcn,Rosner:2013sha}. This validates the treatment here.

\begin{figure}[tb]
    \centering
    \includegraphics[width=\linewidth]{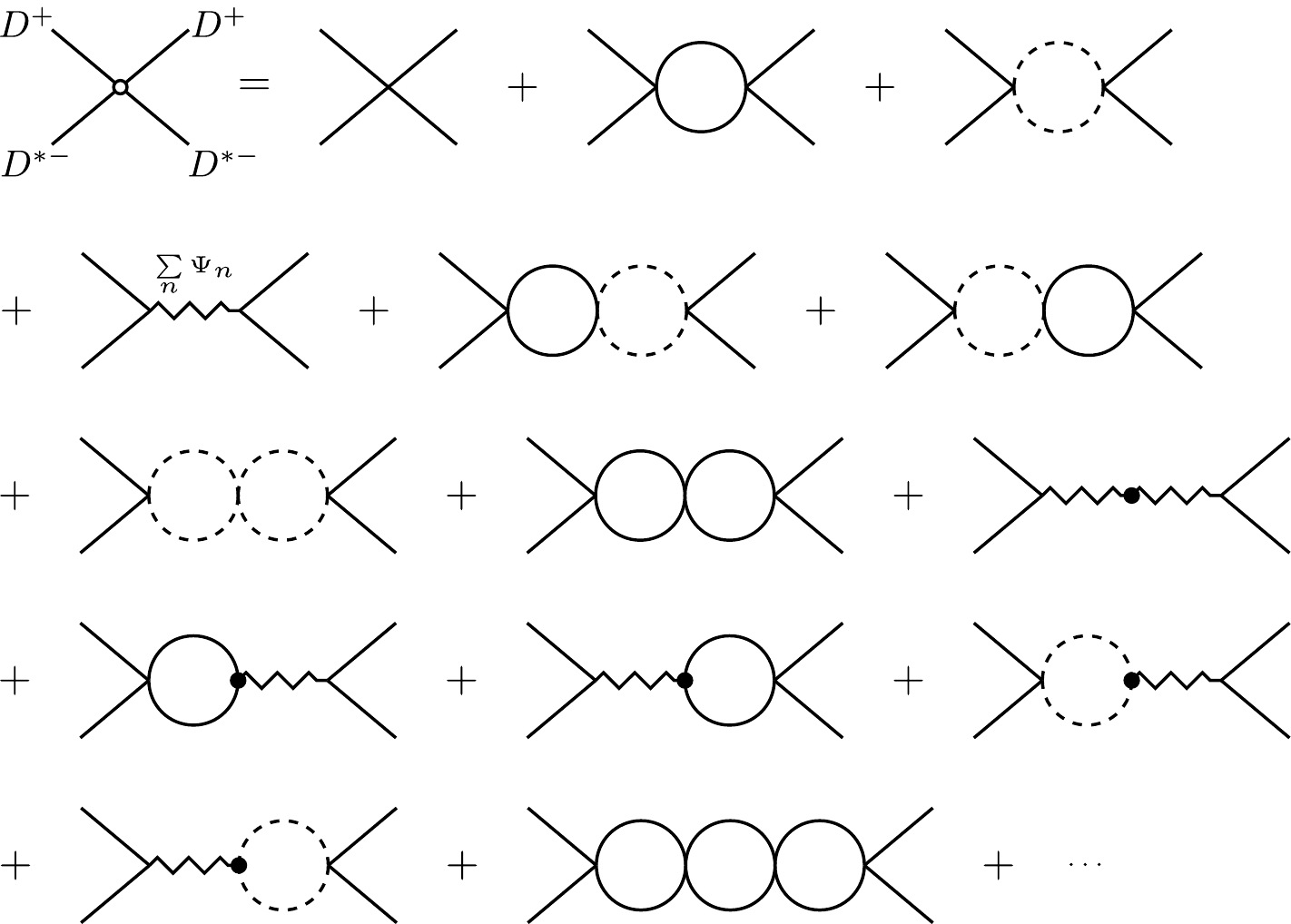}
    \caption{The full amplitude for the $D^+D^{*-}$ near-threshold scattering taking into account both strong interaction and the Coulomb bound states. The solid and dashed lines represent the charged and neutral $D\bar D^*$ states, respectively, and the zigzag lines denote the Coulomb bound states.}
    \label{fig:resummation}
\end{figure}

The $D^+ D^{*-}$ scattering amplitude considering the coupled channels of $D^0\bar D^{*0}$ and $D^+ D^{*-}$, taking into account the $S$-wave Coulomb bound states, can be depicted as Fig.~\ref{fig:resummation}.
The vertex ($S$-matrix element) of the Coulomb bound states coupling to the charmed mesons $D^0\bar D^{*0}$ or $D^+ D^{*-}$ is proportional to the wave function at the origin of the Coulomb bound states,
\begin{equation}
    -i\, C_0 \Psi_{n00}(0) = -i\, C_0 \sqrt{\frac{\alpha^3 \mu_c^3}{\pi n^3}}.
\end{equation}
Then the $S$-wave $T$ matrix for the coupled-channel system with $J^{PC}=1^{++}$ can be written as
\begin{align}
    T(E) = V [1 - G(E)V]^{-1}.
\end{align}
The contact term $V$ and the Green's function matrix $G(E)$ are given by
\begin{align}
    V = C_0 \begin{pmatrix}
        1 & 1  \\ 1 & 1 
    \end{pmatrix}, ~~
    G(E) = \begin{pmatrix}
        J_0(E) & 0  \\
        0 & J_c(E)+J_{|\Psi\rangle}(E)  \\
    \end{pmatrix},
\end{align}
with 
\begin{align}
    J_0(E) &= \frac{\mu_0}{2\pi} \left(-\frac{2\Lambda}{\pi} + \sqrt{-2\mu_0(E + \Delta+ i\Gamma_{0}/2)}\right) ,\notag\\
    J_c(E) &= \frac{\mu_c}{2\pi} \left(-\frac{2\Lambda}{\pi} + \sqrt{-2\mu_c(E+ i\Gamma_{c}/2)}\right), \notag\\
    J_{|\Psi\rangle}(E) &= \sum_{n=1}^\infty \frac{\alpha^3 \mu_c^3}{\pi n^3} \frac1{E + E_n + i\Gamma_c/2}, 
    \label{eq:loop0}
\end{align}
where $E$ is defined as $E= w- \Sigma_c$ with $w$ the center-of-mass energy and $\Sigma_{c(0)}$ the $D^+D^{*-}(D^0\bar D^{*0})$ threshold, $\Delta= \Sigma_c - \Sigma_0$, $\Gamma_{0(c)}$ is the width of the $D^{*0}(D^{*+})$, and $\Lambda$ is a hard cutoff to regularize the ultraviolet (UV) divergence in the loop integrals.
The width of the hadronic atoms that comes from the $D^{*\pm}$ width has been taken into account in $J_{c}(E)$ and $J_{|\Psi\rangle}(E)$.
Here, the $J_c(E)+J_{|\Psi\rangle}(E)$ is a leading order approximation for the exact Coulomb propagator $J_{C}(E)$ in the calculation of ground-state energy eigenvalue, which is given as~\cite{Kong:1999sf} below the $D^{+}D^{*-}$ threshold, 
\begin{align}
   J_{C}(E)=&\,-\frac{\mu_{c} \Lambda}{\pi^{2}}-\frac{\alpha \mu_{c}^{2}}{\pi}\left(\ln \frac{\Lambda}{\alpha \mu_{c}}-\gamma_{E}\right)\\ \notag
    &-\frac{\alpha \mu_{c}^{2}}{\pi}\left[\ln ( \eta)+\frac{1}{2  \eta}-\psi(- \eta)\right]+\order{ \frac{\alpha\mu_c}{\Lambda}},
\end{align}
where $\gamma_E$ is the Euler constant, $\psi(-\eta)$ is the digamma function, and $\eta={\alpha \mu_c}/\sqrt{-2\mu_c(E+i\Gamma_c/2)}$.
The infinitely many Coulomb poles in $J_{|\Psi\rangle}(E)$ appear as the poles of $\psi(-\eta)$ at $\eta=1,2,\ldots$. For the ground-state energy eigenvalue calculation, $J_C(E)$ can be expressed as
\begin{align}
    J_C(E)=J_c(E)+J_{|\Psi\rangle}(E)+\mathcal{O}\left[\alpha\mu_c^2\ln\left(\alpha\sqrt{\frac{\mu_c}{|E|}}\right)\right],
\end{align}
except for the logarithmic divergence which can be absorbed into the contact term. In the following calculations for the ground-state eigenvalue, we neglect the higher order contributions of the scattering states in the exact Coulomb propagator.

The $T$ matrix can be rewritten as
\begin{equation}
    T(E) = \frac{1}{ C_0^{-1} - \left[J_0(E) + J_c(E) + J_{|\Psi\rangle}(E) \right] } \begin{pmatrix}
        1 & 1\\ 1 & 1  
    \end{pmatrix}.
\end{equation}
One is ready to see that the cutoff terms can be absorbed into the contact term by defining $C_{0R}^{-1} = C_0^{-1} + \Lambda(\mu_0+\mu_c)/\pi^2$. 
The $\X$ and the hadronic atoms appear as poles of the $T$ matrix.
The $\X$ pole is located at $E = -\Delta - \delta - i\Gamma_0/2$ where the imaginary part accounts for the finite decay width of the $D^{*0}$~\cite{Artoisenet:2010va} so that the renormalization condition is
\begin{align}
    C_{0R}^{-1} =&\, \frac{\mu_0}{2\pi}\sqrt{2\mu_0\delta} + \frac{\mu_c}{2\pi}\sqrt{2\mu_c \left(\Delta+\delta-i\frac{\delta\Gamma}{2}\right)} \notag\\
    &\, - \sum_{n=1}^\infty \frac{\alpha^3 \mu_c^3}{\pi n^3} \frac1{\Delta + \delta - E_n - i\delta\Gamma/2} \notag\\
    =&\, \frac{\mu_c}{2\pi}\sqrt{2\mu_c \Delta} \left[1 + \order{\frac{\delta}{\Delta},\frac{\delta\Gamma}{\Delta},\frac{\alpha^3\mu_c^{3/2}}{\Delta^{3/2}}}\right],
    \label{eq:c0r}
\end{align}
where $\delta\Gamma = \Gamma_c - \Gamma_0$.
The effective coupling squared in Eq.~\eqref{eq:gx} can then be derived from the residue of $T_{11}(E)$ at the $\X$ pole.

The infinite series of $S$-wave hadronic atoms correspond to poles around the $D^+D^{*-}$ threshold at $-E_{An}- i\Gamma_c/2$:
\begin{align}
    0=&\, C_{0R}^{-1} + i \frac{\mu_0}{2\pi}\sqrt{2\mu_0\left(\Delta-E_{An}-i\frac{\delta\Gamma}{2}\right)} 
    - \frac{\mu_c}{2\pi}\sqrt{2\mu_cE_{An}} \notag\\
    & - \sum_{n=1}^\infty \frac{\alpha^3 \mu_c^3}{\pi n^3} \frac1{-E_{An} + E_n} . 
    \label{eq:atompole}
\end{align} 
The energy level shifts solved from Eq.~\eqref{eq:atompole} can produce the well-known Deser-Goldberger-Baumann-Thirring formula~\cite{Deser:1954vq} at leading order. Applying Eq.~\eqref{eq:c0r}, one obtains the strong interaction correction to the hadronic atom binding energies, $\Delta E_n = E_{An}-E_n$,
\begin{align}
    \Delta E_n = \frac{2\alpha^3\mu_c^2}{n^3\sqrt{2\mu_c\Delta}}\left[-1-i + \order{\alpha\sqrt{\frac{\mu_c}{\Delta}}} \right]^{-1},
\end{align}
where terms suppressed by $\order{\Delta/\mu_c}$ have been neglected.
We get the binding energy of the ground state of the $X$ atom,
\begin{equation}
    \text{Re}\,E_{A1} = E_1 - \frac{\alpha^3\mu_c^2}{\sqrt{2\mu_c\Delta}} \simeq 22.92~\text{keV},
\end{equation}
and the decay width due to the decay into the $D^0\bar D^{*0}-D^{*0}\bar D^0$ channel as well as from the $D^{*}$ width,
\begin{equation}
    \Gamma_c + 2\text{Im}\,E_{A1} = \Gamma_{c} +  \frac{2\alpha^3\mu_c^2}{\sqrt{2\mu_c\Delta}} = (89.2\pm1.8)~\text{keV},
\end{equation}
where the uncertainty comes from that of $\Gamma_c$.
The effective coupling of the hadronic atoms to the charmed meson pair can be computed from the residue of $T_{22}(E)$. For the ground state, we have
\begin{align}
    g_{A1,\text{str}}^2 =&\, \text{lim}_{E\to -E_{A1}-i\Gamma_c/2} (E + E_{A1}+i\Gamma_c/2) T_{22}(E) \notag\\
    =&\, -i \frac{\pi\alpha^3}{\Delta} \left[ 1 + \order{\frac{\alpha^2\mu_c}{\Delta}} \right],
\end{align}
which is dominated by the $n=1$ term in the summation of Eq.~\eqref{eq:atompole} as it should. The couplings for the excited states are suppressed by $1/n^3$ and thus will not be considered.

\begin{figure}[tb]
    \centering
    \includegraphics[width=0.48\linewidth]{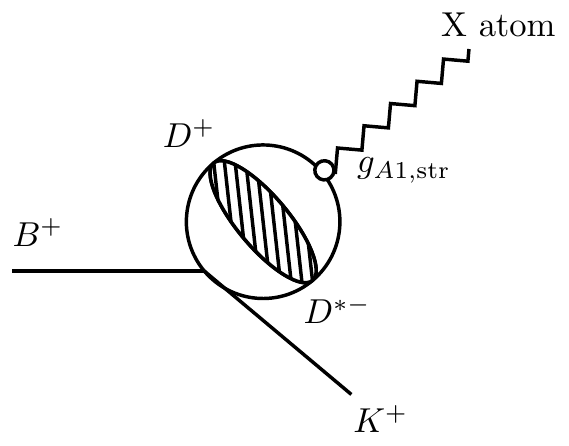}
    \includegraphics[width=0.48\linewidth]{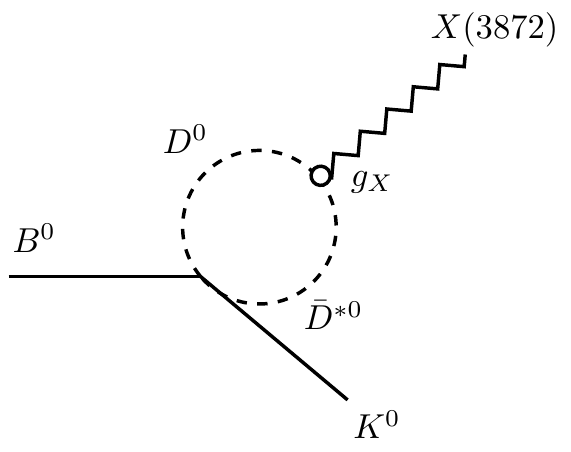}
    \caption{Production of the $X$ atom (left) and $X(3872)$ (right) in $B^+$ and $B^0$ decays. The blob denotes the Coulomb interaction between the charged $D\bar{D}^*$ states.}
    \label{fig:production}
\end{figure}

\begin{figure}[tb]
    \centering
    \includegraphics[width=\linewidth]{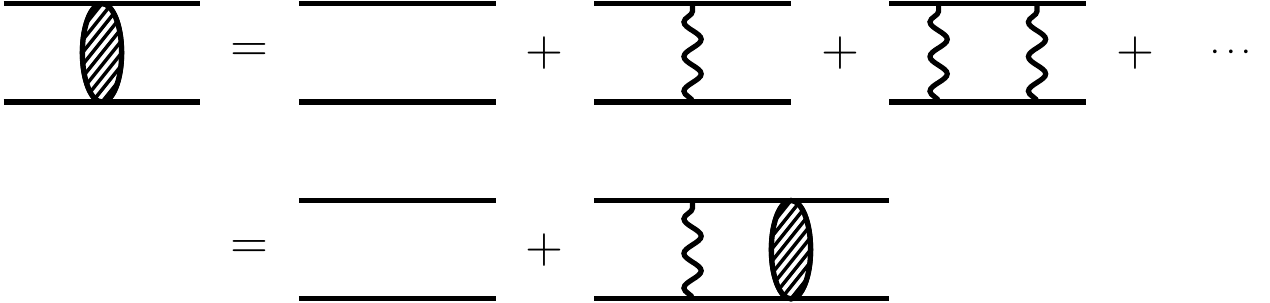}
    \caption{Infinite sum in the Coulomb propagator. The solid lines represent the charged $D\bar{D}^*$ states, and the wavy lines represent the Coulomb photons.}
    \label{fig:Coulomb propagator}
\end{figure}

{\it Production.}---The production of the $X$ atom can be related to that of the $\X$ via isospin symmetry, thus can be used to provide invaluable information on the $\X$ structure. Let us first consider the production in $B$ decays, $B^+ \to (DD^*)_+ K^+ \to A K^+$, $B^0 \to (DD^*)_+^0 K^0 \to X K^0$, where $A$ and $X$ represent the $X$ atom and $\X$, respectively, and $(DD^*)_+$ and $(DD^*)_+^0$ represent the positive $C$-parity pairs of charged and neutral $D\bar D^*$ mesons. 
The processes are depicted in Fig.~\ref{fig:production}, where the Coulomb interaction between the intermediate $D^\pm$ and $D^{*\mp}$ in the $X$ atom production is presented as the Coulomb propagator, an infinite sum of the photon exchange diagrams shown in Fig.~\ref{fig:Coulomb propagator}, and the intermediate $(DD^*)_+$ loop with such Coulomb corrections is given as~\cite{Kong:1999sf}, 
\begin{align}
    G^{\Lambda}_{C}(E)=&\,-\frac{\mu_{c} \Lambda}{\pi^{2}}-\frac{\alpha \mu_{c}^{2}}{\pi}\left(\ln \frac{\Lambda}{\alpha \mu_{c}}-\gamma_{E}\right)\\ \notag
    &-\frac{\alpha \mu_{c}^{2}}{\pi}\left[\ln (i x)+\frac{1}{2 i x}-\psi(-i x)\right]+\order{ \frac{\alpha\mu_c}{\Lambda}},
\end{align}
where $x={\alpha \mu_c}/\sqrt{2\mu_cE}$.
The leading UV divergence is the same as the loop functions in Eq.~\eqref{eq:loop0}; the reason is that the Coulomb exchange as a long-distance effect does not change the UV behavior.

Then the amplitudes can be factorized into a short-distance (SD) piece and a long-distance piece. The latter is given by the effective coupling or universal transition amplitude~\cite{Braaten:2005jj}
\begin{align}
    \A_{B^+\to AK^+} =&\, \A_{B^+\to (DD^*)_+K^+}^\text{SD} g_{A1,\text{str}}, \notag\\
    \A_{B^0\to XK^0} =&\, \A_{B^0\to (DD^*)_+^0K^0}^\text{SD} g_{X},
    \label{eq:B}
\end{align}
where the UV divergence in the loops in Fig.~\ref{fig:production} has been absorbed into the short-distance parts $\A_{B^+\to (DD^*)_+K^+}^\text{SD}$ and $\A_{B^0\to (DD^*)_+^0K^0}^\text{SD}$ through a multiplicative renormalization.
The short-distance factors in these two reactions should be almost the same because of isospin symmetry.
Therefore, we have
\begin{equation}
    R_\Gamma \equiv \frac{\Gamma_{B^+\to AK^+}}{\Gamma_{B^0\to X K^0}} = \frac{|g_{A1,\text{str}}|^2}{ |g_X|^2}. 
    \label{eq:RGamma}
\end{equation}
One can use the nonobservation of the $X$ atom in $B^+$ decays, i.e., an upper bound of $R_\Gamma$, to put a lower bound on the $\X$ binding energy:
\begin{equation}
    \delta \simeq \frac{0.25~\text{eV}}{R_\Gamma^2}.
\end{equation}

Similarly, one can also consider the semi-inclusive production of the $X$ atom and the $\X$ in $pp$ collisions:
$pp\to (DD^*)_+ + y \to A + y$ and $pp\to (DD^*)_+^0 + y \to X + y$, where $y$ denotes the undetected final state particles.
The production amplitudes follow analogous factorization formulas as those in Eq.~\eqref{eq:B}~\cite{Braaten:2019sxh}:
\begin{align}
    \A_{pp\to A+y} =&\, \A_{pp\to (DD^*)_++y}^\text{SD} g_{A1,\text{str}}, \notag\\
    \A_{pp\to X+y} =&\, \A_{pp\to (DD^*)_+^0+y}^\text{SD} g_{X}.
    \label{eq:pp}
\end{align}
At large $p_T$, the production of the charmed meson pairs are dominated by the fragmentation mechanism~\cite{Collins:1989gx}, and the production cross sections for the charged and neutral $D\bar D^*$ pairs are the same in the isospin limit~\cite{Braaten:2001uu}. Therefore, analogous to Eq.~\eqref{eq:RGamma}, we have the following ratio of differential cross sections for the $X$ atom and $\X$ with the same kinematics
\begin{equation}
    R_\sigma \equiv \frac{d\sigma_{pp\to A+y}}{d\sigma_{pp\to X +y}} = \frac{|g_{A1,\text{str}}|^2}{ |g_X|^2},
    \label{eq:Rsigma}
\end{equation}
and 
\begin{equation}
    \delta \simeq \frac{0.25~\text{eV}}{R_\sigma^2}.  
\end{equation}

On the one hand, using the upper limit 180~keV for $\delta$, Eq.~\eqref{eq:delta}, we can predict
\begin{equation}
    R_\Gamma \simeq R_\sigma \gtrsim 1\times 10^{-3}.
\end{equation}
On the other hand, if one can extract an upper bound for $R_\Gamma$ and/or $R_\sigma$, a lower bound for the $\X$ binding energy can be deduced.
Such a lower bound may be obtained from an analysis of the thousands of events collected at the CDF~\cite{Aaltonen:2009vj} and LHCb~\cite{Aaij:2020qga,Aaij:2020xjx} experiments for the $\X$.

{\it Summary.}---To summarize, in this Letter we propose to search for the $X$ atom, which is \iffalse the ground state of\fi the Coulomb bound state of a pair of charged $D\bar D^*$ mesons with positive $C$ parity and a mass of $(3879.89\pm0.07)$~MeV. The width of the $X$ atom comes predominantly from the
width of the $D^{*\pm}$, which is about 80~keV, much larger than the Coulomb binding energy of the $X$ atom, which is about 23~keV. Consequently, the line shape of the $X$ atom will be more like that of toponium~\cite{Sumino:2005gc} than positronium.
There will be a single peak near the $D^{+}D^{*-}$ threshold from all the $S$-wave Coulomb bound states, and there will also be a step in the cross section near that threshold from the production of on-shell charged charm meson pairs.
Moreover, the $X$ atom receives a width due to the decay into the neutral $D\bar D^*$ channels and an energy shift to the Coulomb binding energy due to the strong interaction. 
The values of the width and energy shift are connected to the existence of the $\X$ just around the $D^0\bar D^{*0}$ threshold. A negative $C$-parity $D\bar D^*$ atom should also exist, which, however, cannot be directly connected to the $\X$.
The $X$ atom decays into the same final states as the $\X$. 
However, there is a crucial difference: the $X$ atom couples predominantly to the charged channel so that it couples to both isoscalar and isovector channels with similar strengths, while the $X(3872)$ couplings are approximately of an isoscalar nature~\cite{Gamermann:2009fv,Hanhart:2011tn}.
In addition, its mass is much closer to the $D^+D^{*-}$ threshold than to the $D^0\bar D^{*0}$. 
Then we would expect that 
\begin{equation}
    \frac{\B(A\to J/\psi\pi^+\pi^-)}{\B(A\to J/\psi\pi^+\pi^-\pi^0)} > \frac{\B(X\to J/\psi\pi^+\pi^-)}{\B(X\to J/\psi\pi^+\pi^-\pi^0)}.
\end{equation}

Isospin symmetry allows the production of the $X$ atom to be related to that of the $\X$. From the ratio of the productions of the $X$ atom and the $\X$ in $B$-meson decays and at hadron colliders, one can derive a lower bound on the binding energy of the $\X$. We suggest to search for the $X$ atom, whose production rate relative to the $X(3872)$ should be at least $1\times10^{-3}$, at the LHC upgrades and at PANDA. New insights into the $\X$ mystery are foreseen.

\medskip

\begin{acknowledgments}
    We are grateful to Christoph Hanhart, Yu Jia, Jian-Ping Ma, and Pan-Pan Shi for useful discussions. 
    This work is supported in part by the Chinese Academy of Sciences (CAS) under Grants No.~XDB34030000 and No.~QYZDB-SSW-SYS013, by the National Natural Science Foundation of China (NSFC) under Grants No.~11835015, No.~12047503, and No.~11961141012, by the NSFC and the Deutsche Forschungsgemeinschaft (DFG, German Research Foundation) through the funds provided to the Sino-German Collaborative Research Center TRR110 ``Symmetries and the Emergence of Structure in QCD'' (NSFC Grant No. 12070131001, DFG Project-ID 196253076), and by the CAS Center for Excellence in Particle Physics (CCEPP).
\end{acknowledgments}

\bibliographystyle{apsrev}
\bibliography{xatom}

\begin{thebibliography}{52}
\expandafter\ifx\csname natexlab\endcsname\relax\def\natexlab#1{#1}\fi
\expandafter\ifx\csname bibnamefont\endcsname\relax
  \def\bibnamefont#1{#1}\fi
\expandafter\ifx\csname bibfnamefont\endcsname\relax
  \def\bibfnamefont#1{#1}\fi
\expandafter\ifx\csname citenamefont\endcsname\relax
  \def\citenamefont#1{#1}\fi
\expandafter\ifx\csname url\endcsname\relax
  \def\url#1{\texttt{#1}}\fi
\expandafter\ifx\csname urlprefix\endcsname\relax\def\urlprefix{URL }\fi
\providecommand{\bibinfo}[2]{#2}
\providecommand{\eprint}[2][]{\url{#2}}

\bibitem[{\citenamefont{Ablikim et~al.}(2020)}]{Ablikim:2019hff}
\bibinfo{author}{\bibfnamefont{M.}~\bibnamefont{Ablikim}} \bibnamefont{et~al.}
  (\bibinfo{collaboration}{BESIII}), \bibinfo{journal}{Chin. Phys. C}
  \textbf{\bibinfo{volume}{44}}, \bibinfo{pages}{040001}
  (\bibinfo{year}{2020}), \eprint{1912.05983}.

\bibitem[{\citenamefont{Cerri et~al.}(2019)}]{Cerri:2018ypt}
\bibinfo{author}{\bibfnamefont{A.}~\bibnamefont{Cerri}} \bibnamefont{et~al.},
  \bibinfo{journal}{CERN Yellow Rep. Monogr.} \textbf{\bibinfo{volume}{7}},
  \bibinfo{pages}{867} (\bibinfo{year}{2019}), \eprint{1812.07638}.

\bibitem[{\citenamefont{Altmannshofer et~al.}(2019)}]{Kou:2018nap}
\bibinfo{author}{\bibfnamefont{W.}~\bibnamefont{Altmannshofer}}
  \bibnamefont{et~al.} (\bibinfo{collaboration}{Belle-II}),
  \bibinfo{journal}{PTEP} \textbf{\bibinfo{volume}{2019}},
  \bibinfo{pages}{123C01} (\bibinfo{year}{2019}), \bibinfo{note}{[Erratum: PTEP
  2020, 029201 (2020)]}, \eprint{1808.10567}.

\bibitem[{\citenamefont{Lutz et~al.}(2009)}]{Lutz:2009ff}
\bibinfo{author}{\bibfnamefont{M.}~\bibnamefont{Lutz}} \bibnamefont{et~al.}
  (\bibinfo{collaboration}{PANDA}) (\bibinfo{year}{2009}), \eprint{0903.3905}.

\bibitem[{\citenamefont{Zyla et~al.}(2020)}]{Zyla:2020zbs}
\bibinfo{author}{\bibfnamefont{P.}~\bibnamefont{Zyla}} \bibnamefont{et~al.}
  (\bibinfo{collaboration}{Particle Data Group}), \bibinfo{journal}{PTEP}
  \textbf{\bibinfo{volume}{2020}}, \bibinfo{pages}{083C01}
  (\bibinfo{year}{2020}).

\bibitem[{\citenamefont{Choi et~al.}(2003)}]{Choi:2003ue}
\bibinfo{author}{\bibfnamefont{S.}~\bibnamefont{Choi}} \bibnamefont{et~al.}
  (\bibinfo{collaboration}{Belle}), \bibinfo{journal}{Phys. Rev. Lett.}
  \textbf{\bibinfo{volume}{91}}, \bibinfo{pages}{262001}
  (\bibinfo{year}{2003}), \eprint{hep-ex/0309032}.

\bibitem[{\citenamefont{Chen et~al.}(2016)\citenamefont{Chen, Chen, Liu, and
  Zhu}}]{Chen:2016qju}
\bibinfo{author}{\bibfnamefont{H.-X.} \bibnamefont{Chen}},
  \bibinfo{author}{\bibfnamefont{W.}~\bibnamefont{Chen}},
  \bibinfo{author}{\bibfnamefont{X.}~\bibnamefont{Liu}}, \bibnamefont{and}
  \bibinfo{author}{\bibfnamefont{S.-L.} \bibnamefont{Zhu}},
  \bibinfo{journal}{Phys. Rept.} \textbf{\bibinfo{volume}{639}},
  \bibinfo{pages}{1} (\bibinfo{year}{2016}), \eprint{1601.02092}.

\bibitem[{\citenamefont{Hosaka et~al.}(2016)\citenamefont{Hosaka, Iijima,
  Miyabayashi, Sakai, and Yasui}}]{Hosaka:2016pey}
\bibinfo{author}{\bibfnamefont{A.}~\bibnamefont{Hosaka}},
  \bibinfo{author}{\bibfnamefont{T.}~\bibnamefont{Iijima}},
  \bibinfo{author}{\bibfnamefont{K.}~\bibnamefont{Miyabayashi}},
  \bibinfo{author}{\bibfnamefont{Y.}~\bibnamefont{Sakai}}, \bibnamefont{and}
  \bibinfo{author}{\bibfnamefont{S.}~\bibnamefont{Yasui}},
  \bibinfo{journal}{PTEP} \textbf{\bibinfo{volume}{2016}},
  \bibinfo{pages}{062C01} (\bibinfo{year}{2016}), \eprint{1603.09229}.

\bibitem[{\citenamefont{Lebed et~al.}(2017)\citenamefont{Lebed, Mitchell, and
  Swanson}}]{Lebed:2016hpi}
\bibinfo{author}{\bibfnamefont{R.~F.} \bibnamefont{Lebed}},
  \bibinfo{author}{\bibfnamefont{R.~E.} \bibnamefont{Mitchell}},
  \bibnamefont{and} \bibinfo{author}{\bibfnamefont{E.~S.}
  \bibnamefont{Swanson}}, \bibinfo{journal}{Prog. Part. Nucl. Phys.}
  \textbf{\bibinfo{volume}{93}}, \bibinfo{pages}{143} (\bibinfo{year}{2017}),
  \eprint{1610.04528}.

\bibitem[{\citenamefont{Esposito et~al.}(2017)\citenamefont{Esposito, Pilloni,
  and Polosa}}]{Esposito:2016noz}
\bibinfo{author}{\bibfnamefont{A.}~\bibnamefont{Esposito}},
  \bibinfo{author}{\bibfnamefont{A.}~\bibnamefont{Pilloni}}, \bibnamefont{and}
  \bibinfo{author}{\bibfnamefont{A.}~\bibnamefont{Polosa}},
  \bibinfo{journal}{Phys. Rept.} \textbf{\bibinfo{volume}{668}},
  \bibinfo{pages}{1} (\bibinfo{year}{2017}), \eprint{1611.07920}.

\bibitem[{\citenamefont{Guo et~al.}(2018)\citenamefont{Guo, Hanhart,
  Mei\ss{}ner, Wang, Zhao, and Zou}}]{Guo:2017jvc}
\bibinfo{author}{\bibfnamefont{F.-K.} \bibnamefont{Guo}},
  \bibinfo{author}{\bibfnamefont{C.}~\bibnamefont{Hanhart}},
  \bibinfo{author}{\bibfnamefont{U.-G.} \bibnamefont{Mei\ss{}ner}},
  \bibinfo{author}{\bibfnamefont{Q.}~\bibnamefont{Wang}},
  \bibinfo{author}{\bibfnamefont{Q.}~\bibnamefont{Zhao}}, \bibnamefont{and}
  \bibinfo{author}{\bibfnamefont{B.-S.} \bibnamefont{Zou}},
  \bibinfo{journal}{Rev. Mod. Phys.} \textbf{\bibinfo{volume}{90}},
  \bibinfo{pages}{015004} (\bibinfo{year}{2018}), \eprint{1705.00141}.

\bibitem[{\citenamefont{Ali et~al.}(2017)\citenamefont{Ali, Lange, and
  Stone}}]{Ali:2017jda}
\bibinfo{author}{\bibfnamefont{A.}~\bibnamefont{Ali}},
  \bibinfo{author}{\bibfnamefont{J.~S.} \bibnamefont{Lange}}, \bibnamefont{and}
  \bibinfo{author}{\bibfnamefont{S.}~\bibnamefont{Stone}},
  \bibinfo{journal}{Prog. Part. Nucl. Phys.} \textbf{\bibinfo{volume}{97}},
  \bibinfo{pages}{123} (\bibinfo{year}{2017}), \eprint{1706.00610}.

\bibitem[{\citenamefont{Olsen et~al.}(2018)\citenamefont{Olsen, Skwarnicki, and
  Zieminska}}]{Olsen:2017bmm}
\bibinfo{author}{\bibfnamefont{S.~L.} \bibnamefont{Olsen}},
  \bibinfo{author}{\bibfnamefont{T.}~\bibnamefont{Skwarnicki}},
  \bibnamefont{and}
  \bibinfo{author}{\bibfnamefont{D.}~\bibnamefont{Zieminska}},
  \bibinfo{journal}{Rev. Mod. Phys.} \textbf{\bibinfo{volume}{90}},
  \bibinfo{pages}{015003} (\bibinfo{year}{2018}), \eprint{1708.04012}.

\bibitem[{\citenamefont{Kalashnikova and
  Nefediev}(2019)}]{Kalashnikova:2018vkv}
\bibinfo{author}{\bibfnamefont{Y.~S.} \bibnamefont{Kalashnikova}}
  \bibnamefont{and} \bibinfo{author}{\bibfnamefont{A.}~\bibnamefont{Nefediev}},
  \bibinfo{journal}{Phys. Usp.} \textbf{\bibinfo{volume}{62}},
  \bibinfo{pages}{568} (\bibinfo{year}{2019}), \eprint{1811.01324}.

\bibitem[{\citenamefont{Liu et~al.}(2019)\citenamefont{Liu, Chen, Chen, Liu,
  and Zhu}}]{Liu:2019zoy}
\bibinfo{author}{\bibfnamefont{Y.-R.} \bibnamefont{Liu}},
  \bibinfo{author}{\bibfnamefont{H.-X.} \bibnamefont{Chen}},
  \bibinfo{author}{\bibfnamefont{W.}~\bibnamefont{Chen}},
  \bibinfo{author}{\bibfnamefont{X.}~\bibnamefont{Liu}}, \bibnamefont{and}
  \bibinfo{author}{\bibfnamefont{S.-L.} \bibnamefont{Zhu}},
  \bibinfo{journal}{Prog. Part. Nucl. Phys.} \textbf{\bibinfo{volume}{107}},
  \bibinfo{pages}{237} (\bibinfo{year}{2019}), \eprint{1903.11976}.

\bibitem[{\citenamefont{Brambilla et~al.}(2020)\citenamefont{Brambilla,
  Eidelman, Hanhart, Nefediev, Shen, Thomas, Vairo, and
  Yuan}}]{Brambilla:2019esw}
\bibinfo{author}{\bibfnamefont{N.}~\bibnamefont{Brambilla}},
  \bibinfo{author}{\bibfnamefont{S.}~\bibnamefont{Eidelman}},
  \bibinfo{author}{\bibfnamefont{C.}~\bibnamefont{Hanhart}},
  \bibinfo{author}{\bibfnamefont{A.}~\bibnamefont{Nefediev}},
  \bibinfo{author}{\bibfnamefont{C.-P.} \bibnamefont{Shen}},
  \bibinfo{author}{\bibfnamefont{C.~E.} \bibnamefont{Thomas}},
  \bibinfo{author}{\bibfnamefont{A.}~\bibnamefont{Vairo}}, \bibnamefont{and}
  \bibinfo{author}{\bibfnamefont{C.-Z.} \bibnamefont{Yuan}},
  \bibinfo{journal}{Phys. Rept.} \textbf{\bibinfo{volume}{873}},
  \bibinfo{pages}{1} (\bibinfo{year}{2020}), \eprint{1907.07583}.

\bibitem[{\citenamefont{Aaij et~al.}(2020{\natexlab{a}})}]{Aaij:2020qga}
\bibinfo{author}{\bibfnamefont{R.}~\bibnamefont{Aaij}} \bibnamefont{et~al.}
  (\bibinfo{collaboration}{LHCb}), \bibinfo{journal}{Phys. Rev. D}
  \textbf{\bibinfo{volume}{102}}, \bibinfo{pages}{092005}
  (\bibinfo{year}{2020}{\natexlab{a}}), \eprint{2005.13419}.

\bibitem[{\citenamefont{Aaij et~al.}(2020{\natexlab{b}})}]{Aaij:2020xjx}
\bibinfo{author}{\bibfnamefont{R.}~\bibnamefont{Aaij}} \bibnamefont{et~al.}
  (\bibinfo{collaboration}{LHCb}), \bibinfo{journal}{JHEP}
  \textbf{\bibinfo{volume}{08}}, \bibinfo{pages}{123}
  (\bibinfo{year}{2020}{\natexlab{b}}), \eprint{2005.13422}.

\bibitem[{\citenamefont{Li and Yuan}(2019)}]{Li:2019kpj}
\bibinfo{author}{\bibfnamefont{C.}~\bibnamefont{Li}} \bibnamefont{and}
  \bibinfo{author}{\bibfnamefont{C.-Z.} \bibnamefont{Yuan}},
  \bibinfo{journal}{Phys. Rev. D} \textbf{\bibinfo{volume}{100}},
  \bibinfo{pages}{094003} (\bibinfo{year}{2019}), \eprint{1907.09149}.

\bibitem[{\citenamefont{Braaten
  et~al.}(2019{\natexlab{a}})\citenamefont{Braaten, He, and
  Ingles}}]{Braaten:2019ags}
\bibinfo{author}{\bibfnamefont{E.}~\bibnamefont{Braaten}},
  \bibinfo{author}{\bibfnamefont{L.-P.} \bibnamefont{He}}, \bibnamefont{and}
  \bibinfo{author}{\bibfnamefont{K.}~\bibnamefont{Ingles}}
  (\bibinfo{year}{2019}{\natexlab{a}}), \eprint{1908.02807}.

\bibitem[{\citenamefont{Braaten and Kusunoki}(2005)}]{Braaten:2005jj}
\bibinfo{author}{\bibfnamefont{E.}~\bibnamefont{Braaten}} \bibnamefont{and}
  \bibinfo{author}{\bibfnamefont{M.}~\bibnamefont{Kusunoki}},
  \bibinfo{journal}{Phys. Rev. D} \textbf{\bibinfo{volume}{72}},
  \bibinfo{pages}{014012} (\bibinfo{year}{2005}), \eprint{hep-ph/0506087}.

\bibitem[{\citenamefont{Braaten and Kusunoki}(2004)}]{Braaten:2004rw}
\bibinfo{author}{\bibfnamefont{E.}~\bibnamefont{Braaten}} \bibnamefont{and}
  \bibinfo{author}{\bibfnamefont{M.}~\bibnamefont{Kusunoki}},
  \bibinfo{journal}{Phys. Rev. D} \textbf{\bibinfo{volume}{69}},
  \bibinfo{pages}{114012} (\bibinfo{year}{2004}), \eprint{hep-ph/0402177}.

\bibitem[{\citenamefont{Artoisenet et~al.}(2010)\citenamefont{Artoisenet,
  Braaten, and Kang}}]{Artoisenet:2010va}
\bibinfo{author}{\bibfnamefont{P.}~\bibnamefont{Artoisenet}},
  \bibinfo{author}{\bibfnamefont{E.}~\bibnamefont{Braaten}}, \bibnamefont{and}
  \bibinfo{author}{\bibfnamefont{D.}~\bibnamefont{Kang}},
  \bibinfo{journal}{Phys. Rev. D} \textbf{\bibinfo{volume}{82}},
  \bibinfo{pages}{014013} (\bibinfo{year}{2010}), \eprint{1005.2167}.

\bibitem[{\citenamefont{Guo et~al.}(2014)\citenamefont{Guo, Mei\ss{}ner, Wang,
  and Yang}}]{Guo:2014sca}
\bibinfo{author}{\bibfnamefont{F.-K.} \bibnamefont{Guo}},
  \bibinfo{author}{\bibfnamefont{U.-G.} \bibnamefont{Mei\ss{}ner}},
  \bibinfo{author}{\bibfnamefont{W.}~\bibnamefont{Wang}}, \bibnamefont{and}
  \bibinfo{author}{\bibfnamefont{Z.}~\bibnamefont{Yang}},
  \bibinfo{journal}{Eur. Phys. J. C} \textbf{\bibinfo{volume}{74}},
  \bibinfo{pages}{3063} (\bibinfo{year}{2014}), \eprint{1402.6236}.

\bibitem[{\citenamefont{Bignamini et~al.}(2009)\citenamefont{Bignamini,
  Grinstein, Piccinini, Polosa, and Sabelli}}]{Bignamini:2009sk}
\bibinfo{author}{\bibfnamefont{C.}~\bibnamefont{Bignamini}},
  \bibinfo{author}{\bibfnamefont{B.}~\bibnamefont{Grinstein}},
  \bibinfo{author}{\bibfnamefont{F.}~\bibnamefont{Piccinini}},
  \bibinfo{author}{\bibfnamefont{A.}~\bibnamefont{Polosa}}, \bibnamefont{and}
  \bibinfo{author}{\bibfnamefont{C.}~\bibnamefont{Sabelli}},
  \bibinfo{journal}{Phys. Rev. Lett.} \textbf{\bibinfo{volume}{103}},
  \bibinfo{pages}{162001} (\bibinfo{year}{2009}), \eprint{0906.0882}.

\bibitem[{\citenamefont{Artoisenet and Braaten}(2010)}]{Artoisenet:2009wk}
\bibinfo{author}{\bibfnamefont{P.}~\bibnamefont{Artoisenet}} \bibnamefont{and}
  \bibinfo{author}{\bibfnamefont{E.}~\bibnamefont{Braaten}},
  \bibinfo{journal}{Phys. Rev. D} \textbf{\bibinfo{volume}{81}},
  \bibinfo{pages}{114018} (\bibinfo{year}{2010}), \eprint{0911.2016}.

\bibitem[{\citenamefont{Meng et~al.}(2017)\citenamefont{Meng, Han, and
  Chao}}]{Meng:2013gga}
\bibinfo{author}{\bibfnamefont{C.}~\bibnamefont{Meng}},
  \bibinfo{author}{\bibfnamefont{H.}~\bibnamefont{Han}}, \bibnamefont{and}
  \bibinfo{author}{\bibfnamefont{K.-T.} \bibnamefont{Chao}},
  \bibinfo{journal}{Phys. Rev. D} \textbf{\bibinfo{volume}{96}},
  \bibinfo{pages}{074014} (\bibinfo{year}{2017}), \eprint{1304.6710}.

\bibitem[{\citenamefont{Albaladejo et~al.}(2017)\citenamefont{Albaladejo, Guo,
  Hanhart, Mei\ss{}ner, Nieves, Nogga, and Yang}}]{Albaladejo:2017blx}
\bibinfo{author}{\bibfnamefont{M.}~\bibnamefont{Albaladejo}},
  \bibinfo{author}{\bibfnamefont{F.-K.} \bibnamefont{Guo}},
  \bibinfo{author}{\bibfnamefont{C.}~\bibnamefont{Hanhart}},
  \bibinfo{author}{\bibfnamefont{U.-G.} \bibnamefont{Mei\ss{}ner}},
  \bibinfo{author}{\bibfnamefont{J.}~\bibnamefont{Nieves}},
  \bibinfo{author}{\bibfnamefont{A.}~\bibnamefont{Nogga}}, \bibnamefont{and}
  \bibinfo{author}{\bibfnamefont{Z.}~\bibnamefont{Yang}},
  \bibinfo{journal}{Chin. Phys. C} \textbf{\bibinfo{volume}{41}},
  \bibinfo{pages}{121001} (\bibinfo{year}{2017}), \eprint{1709.09101}.

\bibitem[{\citenamefont{Esposito et~al.}(2018)\citenamefont{Esposito,
  Grinstein, Maiani, Piccinini, Pilloni, Polosa, and
  Riquer}}]{Esposito:2017qef}
\bibinfo{author}{\bibfnamefont{A.}~\bibnamefont{Esposito}},
  \bibinfo{author}{\bibfnamefont{B.}~\bibnamefont{Grinstein}},
  \bibinfo{author}{\bibfnamefont{L.}~\bibnamefont{Maiani}},
  \bibinfo{author}{\bibfnamefont{F.}~\bibnamefont{Piccinini}},
  \bibinfo{author}{\bibfnamefont{A.}~\bibnamefont{Pilloni}},
  \bibinfo{author}{\bibfnamefont{A.}~\bibnamefont{Polosa}}, \bibnamefont{and}
  \bibinfo{author}{\bibfnamefont{V.}~\bibnamefont{Riquer}},
  \bibinfo{journal}{Chin. Phys. C} \textbf{\bibinfo{volume}{42}},
  \bibinfo{pages}{114107} (\bibinfo{year}{2018}), \eprint{1709.09631}.

\bibitem[{\citenamefont{Wang}(2018)}]{Wang:2017gay}
\bibinfo{author}{\bibfnamefont{W.}~\bibnamefont{Wang}}, \bibinfo{journal}{Chin.
  Phys. C} \textbf{\bibinfo{volume}{42}}, \bibinfo{pages}{043103}
  (\bibinfo{year}{2018}), \eprint{1709.10382}.

\bibitem[{\citenamefont{Braaten
  et~al.}(2019{\natexlab{b}})\citenamefont{Braaten, He, and
  Ingles}}]{Braaten:2018eov}
\bibinfo{author}{\bibfnamefont{E.}~\bibnamefont{Braaten}},
  \bibinfo{author}{\bibfnamefont{L.-P.} \bibnamefont{He}}, \bibnamefont{and}
  \bibinfo{author}{\bibfnamefont{K.}~\bibnamefont{Ingles}},
  \bibinfo{journal}{Phys. Rev. D} \textbf{\bibinfo{volume}{100}},
  \bibinfo{pages}{094024} (\bibinfo{year}{2019}{\natexlab{b}}),
  \eprint{1811.08876}.

\bibitem[{\citenamefont{Butenschoen et~al.}(2019)\citenamefont{Butenschoen, He,
  and Kniehl}}]{Butenschoen:2019npa}
\bibinfo{author}{\bibfnamefont{M.}~\bibnamefont{Butenschoen}},
  \bibinfo{author}{\bibfnamefont{Z.-G.} \bibnamefont{He}}, \bibnamefont{and}
  \bibinfo{author}{\bibfnamefont{B.~A.} \bibnamefont{Kniehl}},
  \bibinfo{journal}{Phys. Rev. Lett.} \textbf{\bibinfo{volume}{123}},
  \bibinfo{pages}{032001} (\bibinfo{year}{2019}), \eprint{1906.08553}.

\bibitem[{\citenamefont{Zhang et~al.}(2021)\citenamefont{Zhang, Liao, Wang,
  Wang, and Xing}}]{Zhang:2020dwn}
\bibinfo{author}{\bibfnamefont{H.}~\bibnamefont{Zhang}},
  \bibinfo{author}{\bibfnamefont{J.}~\bibnamefont{Liao}},
  \bibinfo{author}{\bibfnamefont{E.}~\bibnamefont{Wang}},
  \bibinfo{author}{\bibfnamefont{Q.}~\bibnamefont{Wang}}, \bibnamefont{and}
  \bibinfo{author}{\bibfnamefont{H.}~\bibnamefont{Xing}},
  \bibinfo{journal}{Phys. Rev. Lett.} \textbf{\bibinfo{volume}{126}},
  \bibinfo{pages}{012301} (\bibinfo{year}{2021}), \eprint{2004.00024}.

\bibitem[{\citenamefont{Esposito et~al.}(2020)\citenamefont{Esposito, Ferreiro,
  Pilloni, Polosa, and Salgado}}]{Esposito:2020ywk}
\bibinfo{author}{\bibfnamefont{A.}~\bibnamefont{Esposito}},
  \bibinfo{author}{\bibfnamefont{E.~G.} \bibnamefont{Ferreiro}},
  \bibinfo{author}{\bibfnamefont{A.}~\bibnamefont{Pilloni}},
  \bibinfo{author}{\bibfnamefont{A.~D.} \bibnamefont{Polosa}},
  \bibnamefont{and} \bibinfo{author}{\bibfnamefont{C.~A.}
  \bibnamefont{Salgado}} (\bibinfo{year}{2020}), \eprint{2006.15044}.

\bibitem[{\citenamefont{Barucca et~al.}(2019)}]{PANDA:2018zjt}
\bibinfo{author}{\bibfnamefont{G.}~\bibnamefont{Barucca}} \bibnamefont{et~al.}
  (\bibinfo{collaboration}{PANDA}), \bibinfo{journal}{Eur. Phys. J. A}
  \textbf{\bibinfo{volume}{55}}, \bibinfo{pages}{42} (\bibinfo{year}{2019}),
  \eprint{1812.05132}.

\bibitem[{\citenamefont{Sakai et~al.}(2020)\citenamefont{Sakai, Jing, and
  Guo}}]{Sakai:2020crh}
\bibinfo{author}{\bibfnamefont{S.}~\bibnamefont{Sakai}},
  \bibinfo{author}{\bibfnamefont{H.-J.} \bibnamefont{Jing}}, \bibnamefont{and}
  \bibinfo{author}{\bibfnamefont{F.-K.} \bibnamefont{Guo}},
  \bibinfo{journal}{Phys. Rev. D} \textbf{\bibinfo{volume}{102}},
  \bibinfo{pages}{114041} (\bibinfo{year}{2020}), \eprint{2008.10829}.

\bibitem[{\citenamefont{Guo}(2019)}]{Guo:2019qcn}
\bibinfo{author}{\bibfnamefont{F.-K.} \bibnamefont{Guo}},
  \bibinfo{journal}{Phys. Rev. Lett.} \textbf{\bibinfo{volume}{122}},
  \bibinfo{pages}{202002} (\bibinfo{year}{2019}), \eprint{1902.11221}.

\bibitem[{\citenamefont{Gasser et~al.}(2008)\citenamefont{Gasser, Lyubovitskij,
  and Rusetsky}}]{Gasser:2007zt}
\bibinfo{author}{\bibfnamefont{J.}~\bibnamefont{Gasser}},
  \bibinfo{author}{\bibfnamefont{V.}~\bibnamefont{Lyubovitskij}},
  \bibnamefont{and} \bibinfo{author}{\bibfnamefont{A.}~\bibnamefont{Rusetsky}},
  \bibinfo{journal}{Phys. Rept.} \textbf{\bibinfo{volume}{456}},
  \bibinfo{pages}{167} (\bibinfo{year}{2008}), \eprint{0711.3522}.

\bibitem[{\citenamefont{Gasser et~al.}(2009)\citenamefont{Gasser, Lyubovitskij,
  and Rusetsky}}]{Gasser:2009wf}
\bibinfo{author}{\bibfnamefont{J.}~\bibnamefont{Gasser}},
  \bibinfo{author}{\bibfnamefont{V.}~\bibnamefont{Lyubovitskij}},
  \bibnamefont{and} \bibinfo{author}{\bibfnamefont{A.}~\bibnamefont{Rusetsky}},
  \bibinfo{journal}{Ann. Rev. Nucl. Part. Sci.} \textbf{\bibinfo{volume}{59}},
  \bibinfo{pages}{169} (\bibinfo{year}{2009}), \eprint{0903.0257}.

\bibitem[{\citenamefont{Gamermann et~al.}(2010)\citenamefont{Gamermann, Nieves,
  Oset, and Ruiz~Arriola}}]{Gamermann:2009uq}
\bibinfo{author}{\bibfnamefont{D.}~\bibnamefont{Gamermann}},
  \bibinfo{author}{\bibfnamefont{J.}~\bibnamefont{Nieves}},
  \bibinfo{author}{\bibfnamefont{E.}~\bibnamefont{Oset}}, \bibnamefont{and}
  \bibinfo{author}{\bibfnamefont{E.}~\bibnamefont{Ruiz~Arriola}},
  \bibinfo{journal}{Phys. Rev. D} \textbf{\bibinfo{volume}{81}},
  \bibinfo{pages}{014029} (\bibinfo{year}{2010}), \eprint{0911.4407}.

\bibitem[{\citenamefont{Hanhart et~al.}(2012)\citenamefont{Hanhart,
  Kalashnikova, Kudryavtsev, and Nefediev}}]{Hanhart:2011tn}
\bibinfo{author}{\bibfnamefont{C.}~\bibnamefont{Hanhart}},
  \bibinfo{author}{\bibfnamefont{Y.~S.} \bibnamefont{Kalashnikova}},
  \bibinfo{author}{\bibfnamefont{A.~E.} \bibnamefont{Kudryavtsev}},
  \bibnamefont{and} \bibinfo{author}{\bibfnamefont{A.~V.}
  \bibnamefont{Nefediev}}, \bibinfo{journal}{Phys. Rev. D}
  \textbf{\bibinfo{volume}{85}}, \bibinfo{pages}{011501}
  (\bibinfo{year}{2012}), \eprint{1111.6241}.

\bibitem[{\citenamefont{Braaten et~al.}(2021)\citenamefont{Braaten, He, and
  Jiang}}]{Braaten:2020nmc}
\bibinfo{author}{\bibfnamefont{E.}~\bibnamefont{Braaten}},
  \bibinfo{author}{\bibfnamefont{L.-P.} \bibnamefont{He}}, \bibnamefont{and}
  \bibinfo{author}{\bibfnamefont{J.}~\bibnamefont{Jiang}},
  \bibinfo{journal}{Phys. Rev. D} \textbf{\bibinfo{volume}{103}},
  \bibinfo{pages}{036014} (\bibinfo{year}{2021}), \eprint{2010.05801}.

\bibitem[{\citenamefont{Hanhart et~al.}(2010)\citenamefont{Hanhart,
  Kalashnikova, and Nefediev}}]{Hanhart:2010wh}
\bibinfo{author}{\bibfnamefont{C.}~\bibnamefont{Hanhart}},
  \bibinfo{author}{\bibfnamefont{Y.~S.} \bibnamefont{Kalashnikova}},
  \bibnamefont{and} \bibinfo{author}{\bibfnamefont{A.~V.}
  \bibnamefont{Nefediev}}, \bibinfo{journal}{Phys. Rev. D}
  \textbf{\bibinfo{volume}{81}}, \bibinfo{pages}{094028}
  (\bibinfo{year}{2010}), \eprint{1002.4097}.

\bibitem[{\citenamefont{Rosner}(2013)}]{Rosner:2013sha}
\bibinfo{author}{\bibfnamefont{J.~L.} \bibnamefont{Rosner}},
  \bibinfo{journal}{Phys. Rev. D} \textbf{\bibinfo{volume}{88}},
  \bibinfo{pages}{034034} (\bibinfo{year}{2013}), \eprint{1307.2550}.

\bibitem[{\citenamefont{Kong and Ravndal}(2000)}]{Kong:1999sf}
\bibinfo{author}{\bibfnamefont{X.}~\bibnamefont{Kong}} \bibnamefont{and}
  \bibinfo{author}{\bibfnamefont{F.}~\bibnamefont{Ravndal}},
  \bibinfo{journal}{Nucl. Phys. A} \textbf{\bibinfo{volume}{665}},
  \bibinfo{pages}{137} (\bibinfo{year}{2000}), \eprint{hep-ph/9903523}.

\bibitem[{\citenamefont{Deser et~al.}(1954)\citenamefont{Deser, Goldberger,
  Baumann, and Thirring}}]{Deser:1954vq}
\bibinfo{author}{\bibfnamefont{S.}~\bibnamefont{Deser}},
  \bibinfo{author}{\bibfnamefont{M.~L.} \bibnamefont{Goldberger}},
  \bibinfo{author}{\bibfnamefont{K.}~\bibnamefont{Baumann}}, \bibnamefont{and}
  \bibinfo{author}{\bibfnamefont{W.~E.} \bibnamefont{Thirring}},
  \bibinfo{journal}{Phys. Rev.} \textbf{\bibinfo{volume}{96}},
  \bibinfo{pages}{774} (\bibinfo{year}{1954}).

\bibitem[{\citenamefont{Braaten
  et~al.}(2019{\natexlab{c}})\citenamefont{Braaten, He, and
  Ingles}}]{Braaten:2019sxh}
\bibinfo{author}{\bibfnamefont{E.}~\bibnamefont{Braaten}},
  \bibinfo{author}{\bibfnamefont{L.-P.} \bibnamefont{He}}, \bibnamefont{and}
  \bibinfo{author}{\bibfnamefont{K.}~\bibnamefont{Ingles}},
  \bibinfo{journal}{Phys. Rev. D} \textbf{\bibinfo{volume}{100}},
  \bibinfo{pages}{094006} (\bibinfo{year}{2019}{\natexlab{c}}),
  \eprint{1903.04355}.

\bibitem[{\citenamefont{Collins et~al.}(1989)\citenamefont{Collins, Soper, and
  Sterman}}]{Collins:1989gx}
\bibinfo{author}{\bibfnamefont{J.~C.} \bibnamefont{Collins}},
  \bibinfo{author}{\bibfnamefont{D.~E.} \bibnamefont{Soper}}, \bibnamefont{and}
  \bibinfo{author}{\bibfnamefont{G.~F.} \bibnamefont{Sterman}},
  \bibinfo{journal}{Adv. Ser. Direct. High Energy Phys.}
  \textbf{\bibinfo{volume}{5}}, \bibinfo{pages}{1} (\bibinfo{year}{1989}),
  \eprint{hep-ph/0409313}.

\bibitem[{\citenamefont{Braaten et~al.}(2002)\citenamefont{Braaten, Jia, and
  Mehen}}]{Braaten:2001uu}
\bibinfo{author}{\bibfnamefont{E.}~\bibnamefont{Braaten}},
  \bibinfo{author}{\bibfnamefont{Y.}~\bibnamefont{Jia}}, \bibnamefont{and}
  \bibinfo{author}{\bibfnamefont{T.}~\bibnamefont{Mehen}},
  \bibinfo{journal}{Phys. Rev. D} \textbf{\bibinfo{volume}{66}},
  \bibinfo{pages}{014003} (\bibinfo{year}{2002}), \eprint{hep-ph/0111296}.

\bibitem[{\citenamefont{Aaltonen et~al.}(2009)}]{Aaltonen:2009vj}
\bibinfo{author}{\bibfnamefont{T.}~\bibnamefont{Aaltonen}} \bibnamefont{et~al.}
  (\bibinfo{collaboration}{CDF}), \bibinfo{journal}{Phys. Rev. Lett.}
  \textbf{\bibinfo{volume}{103}}, \bibinfo{pages}{152001}
  (\bibinfo{year}{2009}), \eprint{0906.5218}.

\bibitem[{\citenamefont{Sumino}(2005)}]{Sumino:2005gc}
\bibinfo{author}{\bibfnamefont{Y.}~\bibnamefont{Sumino}},
  \bibinfo{journal}{Adv. Ser. Direct. High Energy Phys.}
  \textbf{\bibinfo{volume}{19}}, \bibinfo{pages}{135} (\bibinfo{year}{2005}).

\bibitem[{\citenamefont{Gamermann and Oset}(2009)}]{Gamermann:2009fv}
\bibinfo{author}{\bibfnamefont{D.}~\bibnamefont{Gamermann}} \bibnamefont{and}
  \bibinfo{author}{\bibfnamefont{E.}~\bibnamefont{Oset}},
  \bibinfo{journal}{Phys. Rev. D} \textbf{\bibinfo{volume}{80}},
  \bibinfo{pages}{014003} (\bibinfo{year}{2009}), \eprint{0905.0402}.

\end{thebibliography}

\end{document}